# Angular anisotropy of secondary neutron spectra in $^{232}$Th+n

V. M. Maslov[1]

*220025 Minsk, Byelorussia*

Neutron emission spectra (NES) of $^{232}$Th+$n$ interaction provide strong evidence of angular anisotropy of secondary neutron emission, another evidence might be predicted in $^{232}$Th($n,F$) prompt fission neutron spectra (PFNS). In case of NES observed angular anisotropy is presumably due to angular dependence of elastic scattering, direct excitation of collective levels and pre-equilibrium emission of $(n,nX)^1$ neutrons. In $^{232}$Th+$n$ direct excitation data analysis, ground state band levels $J^\pi = 0^+, 2^+, 4^+, 6^+, 8^+$ are coupled within rigid rotator model, while those of $\beta$–bands with $K^\pi = 0^+$, $\gamma$–bands with $K^\pi = 2^+$ and octupole band $K^\pi = 0^-$ are coupled within soft deformable rotator model. NES of $^{232}$Th+$n$ at $E_n \sim 6, \sim 12, \sim 14, \sim 18$ MeV exhaustively described. The net effect of these procedures for $E_n$ up to ~20 MeV is the adequate approximation of angular distributions of $^{232}$Th($n,nX$)$^1$ first neutron inelastic scattering in continuum, which corresponds to $U= 1.2$~6 MeV excitations of $^{232}$Th.

The contribution of $^{232}$Th($n,F$) PFNS to the NES is exceptionally low. PFNS anisotropy occurs because some portion of $(n,nX)^1$ neutrons might be involved in exclusive pre-fission neutron spectra. In $^{232}$Th($n,xnf$) reactions PFNS demonstrate different response to forward and backward $(n,xnf)^1$ neutron emission relative to the incident neutron momentum, when compared with $^{235}$U($n,xnf$) or $^{239}$Pu($n,xnf$) reactions. Average energy of $(n,xnf)^1$ neutrons depends on the neutron emission angle $\theta$, i.e. fission cross section, prompt neutron number and total kinetic energy are shown to vary with the angle $\theta$ as well. Exclusive neutron spectra $(n,xnf)^{1...x}$ at $\theta \sim 90°$ are consistent with observed $^{232}$Th($n,F$) and $^{232}$Th($n,xn$) reaction cross sections within $E_n \sim 1$–20 MeV energy range. Exclusive neutron spectra of $(n,xnf)^{1...x}$, $(n,n\gamma)$ and $(n,xn)^{1...x}$ reactions are calculated with Hauser-Feshbach formalism simultaneously with $(n,F)$ and $(n,xn)$ reaction cross sections, angular dependence of first neutron $(n,nX)^1$ emission $\omega(\theta)$ being included.

Neutron emission spectra (NES) of $^{232}$Th+$n$ interaction provide strong evidence of angular anisotropy of secondary neutron spectra [1], as observed in [2, 3]. Another evidence might be predicted in $^{232}$Th($n,F$) prompt fission neutron spectra (PFNS) anisotropy in a similar fashion as for $^{238}$U($n,F$) [4]. In case of NES observed angular anisotropy of neutron emission is mostly due to angular dependence of elastic scattering, direct excitation of collective levels and pre-equilibrium emission of $(n,nX)^1$ neutrons [5, 6]. In $^{232}$Th+$n$ direct excitation data analysis, ground state band levels $J^\pi = 0^+, 2^+, 4^+, 6^+, 8^+$ are coupled within rigid rotator model, while those of $\beta$–bands with $K^\pi = 0^+$ and $\gamma$–bands with $K^\pi = 2^+$ and octupole band $K^\pi = 0^-$ are coupled within soft deformable rotator model [7, 8]. Levels of second $K^\pi = 0^+$ band (0.73035 MeV) classified as quadrupole longitudinal $\beta$-vibrations, while levels of third $K^\pi = 0^+$ band (1.0787 MeV) - as quadrupole transversal $\gamma$-vibrations. Both defined by softness parameters to respective vibrations $\mu_\beta$ and $\mu_\gamma$ [1, 7–9]. Anomalous rotational $\gamma$-band $K^\pi = 2^+$ levels characterized by the non-axiality parameter $\gamma_o$. That latter band lies much lower (~0.3 MeV) than respective band in case of $^{238}$U, at the other hand this band lowering is accompanied by shift of $K^\pi = 0^+$ band (1.0787 MeV) -that of quadrupole transversal $\gamma$-vibrations, to higher excitation (by ~0.250 MeV), than in case of $^{238}$U nuclide. Quadrupole longitudinal $\beta$-vibration band levels also lowered as compared with relevant band of $^{238}$U nuclide (by ~0.250 MeV). That means $^{232}$Th nuclide, within soft rotator model, is much softer with respect to quadrupole longitudinal $\beta$-vibrations, which pronounces as higher $\mu_\beta$ parameter values [7, 8]. As regards quadrupole transversal $\gamma$-vibrations, static non-axiality parameter $\gamma_o$ for $^{232}$Th is higher than in case of $^{238}$U. Its value was extracted by fitting position of band-head of anomalous $\gamma$–band $K^\pi = 2^+$, which is appreciably lower than in case of $^{238}$U. This mask possible difference of softness to transversal $\gamma$-vibrations of $^{232}$Th and $^{238}$U, i.e. $\mu_\gamma$ parameter values differ only slightly [1, 7–9]. In $^{232}$Th+$n$ direct level excitation data analysis [1, 7–9] ground state band levels $J^\pi = 0^+, 2^+, 4^+, 6^+, 8^+$ are coupled within rigid rotator model, while those of $\beta$-bands with $K^\pi = 0^+$, $\gamma$–bands with $K^\pi = 2^+$ and octupole band $K^\pi = 0^-$ are modelled within soft deformable rotator model ($^{232}$Th levels excitation energies $U$=0~1.2 MeV). Actually, the calculation of direct inelastic scattering [1, 7–9] was made adding each of 17 levels of $K^\pi = 0^+, 2^+, 0^-$ bands, one by one, to the $0^+$–$2^+$–$4^+$–$6^+$–$8^+$ coupling basis, replacing the last $8^+$ member of ground state rotational band. That is justified, since the coupling with ground state band levels is the strongest for any band level. This procedure only slightly changes total and reaction cross sections, the same response is to the increase of coupling basis, i.e. from 3 to 5 levels within a rigid rotator model. At $E_n < E_{nnf}$, $E_{nnf}$ being the threshold of $^{232}$Th($n,nf$) reaction, $^{232}$Th+$n$ NES in the vicinity of the elastic peak are analyzed in [1], the same procedure [7–9] as in case of $^{238}$U+$n$ was applied.

---

[1] mvmmvm1955@mail.ru

In the energy range $E_{nnf} < E_n < 20$ MeV double differential NES are a superposition of prompt fission neutron spectra $S(\varepsilon, E_n, \theta)$, normalized to unity exclusive spectra of $(n,n\gamma)^1$, $(n,2n)^{1,2}$ и $(n,3n)^{1,2,3}$, $\dfrac{d^2\sigma_{nxn}^k(\varepsilon, E_n, \theta)}{d\varepsilon d\theta}$ and spectra of elastic and inelastic scattered neutrons, followed by excitation of collective levels $(n, n')_d$ of $^{232}$Th, $\dfrac{d^2\sigma_{nn\gamma}(\varepsilon, E_q, E_n, \theta)}{d\varepsilon d\theta}$:

$$\begin{aligned}\dfrac{d^2\sigma(\varepsilon, E_n, \theta)}{d\varepsilon d\theta} = \dfrac{1}{2\pi}\Bigg[&v_p(E_n,\theta)\sigma_{nF}(E_n,\theta)S(\varepsilon,E_n,\theta) + \sigma_{nn\gamma}(\varepsilon,E_n,\theta)\dfrac{d^2\sigma_{nn\gamma}^1(\varepsilon,E_n,\theta)}{d\varepsilon d\theta} + \\ &\sigma_{n2n}(\varepsilon,E_n,\theta)\left(\dfrac{d^2\sigma_{n2n}^1(\varepsilon,E_n,\theta)}{d\varepsilon d\theta} + \dfrac{d^2\sigma_{n2n}^2(\varepsilon,E_n,\theta)}{d\varepsilon d\theta}\right) + \\ &\sigma_{n3n}(\varepsilon,E_n,\theta)\left(\dfrac{d^2\sigma_{n3n}^1(\varepsilon,E_n,\theta)}{d\varepsilon d\theta} + \dfrac{d^2\sigma_{n3n}^2(\varepsilon,E_n,\theta)}{d\varepsilon d\theta} + \dfrac{d^2\sigma_{n3n}^3(\varepsilon,E_n,\theta)}{d\varepsilon d\theta}\right) + \\ &\sum_q \dfrac{d\sigma_{nn\gamma}(\varepsilon,E_q,E_n,\theta)}{d\theta}G(\varepsilon,E_q,E_n,\Delta_\theta)\Bigg],\end{aligned} \quad (1)$$

$$G(\varepsilon, E_q, E_n, \Delta_\theta) = \dfrac{2}{\Delta_\theta \sqrt{\pi}}\exp\left\{-\left[\dfrac{\varepsilon - (E_n - E_q)}{\Delta_\theta}\right]^2\right\}. \quad (2)$$

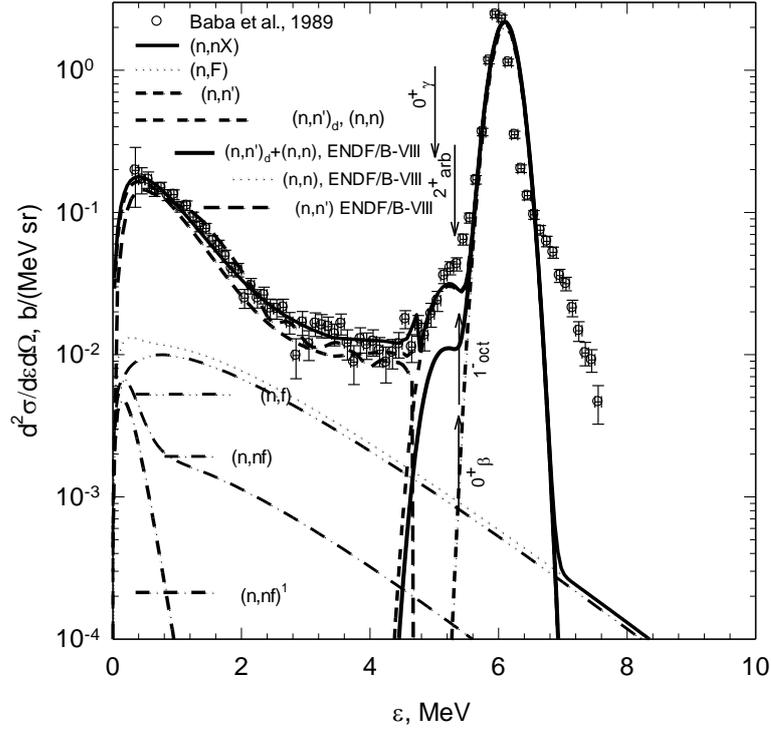

Fig. 1 Double differential NES of $^{232}$Th+$n$ for $E_n = 6.1$ MeV, $\theta \approx 30°$, and its partial constituents; full line – $(n,nX)$; dotted line – $(n,F)$; dashed line – $(n,n\gamma)^1$; dash double dotted line – $n,2n)^1$; dashed line – $(n,2n)^2$; dash–dotted line – $(n,3n)^1$; dashed line – $(n,3n)^2$; full line – $(n,3n)^3$; dashed line – $(n,n)_d+(n,n\gamma)$ for discrete levels; ○ – [2, 3].

In equation (2) $G(\varepsilon, E_q, E_n, \Delta_\theta)$ –resolution function, which depends on $E_n$ and only weakly depends on angle $\theta$. The NES are normalized with average prompt fission neutron number, $(n,xn)$ and $(n,F)$ cross section values.

The QRPA [10] methods are the most advanced in the field, however they still incapable to describe NES of heavy nuclides like $^{238}$U or $^{232}$Th, when inelastic scattering in continuum corresponds to the excitations of residual nuclei $U= 1.2$~$6$ MeV. Emission spectrum of $(n,nX)^1$ reaction, $\dfrac{d^2\sigma^1_{nnx}(\varepsilon, E_n, \theta)}{d\varepsilon d\theta}$, could be represented by the sum of compound and pre-equilibrium components, both weakly dependent on emission angle, and phenomenological function, modelling energy and angle dependence of NES [2, 3]. Double differential spectra of first neutron inelastic scattering in continuum is approximated as

$$\frac{d^2\sigma^1_{nnx}(\varepsilon, E_n, \theta)}{d\varepsilon d\theta} \approx \frac{d^2\tilde{\sigma}^1_{nnx}(\varepsilon, E_n, \theta)}{d\varepsilon d\theta} + \sqrt{\frac{\varepsilon}{E_n}} \frac{\omega(\theta)}{E_n - \varepsilon} \quad (3)$$

$$\omega(\theta) = 0.4\cos^3(\theta) + 0.16 \quad (4)$$

The value of the second term at right hand side of equation (3) depends on the lumped contribution of the direct excitation of the collective levels of $\beta$–bands with $K^\pi = 0^+$, $\gamma$–bands $K^\pi =0^+$, $\gamma$ –bands with $K^\pi =2^+$ and octupole band $K^\pi = 0^-$. Angle-averaged function $\omega(\theta)$, $\langle\omega(\theta)\rangle_\theta$ for scattering angles $\theta_2 - \theta_1 = 135° - 30°$, approximated as $\langle\omega(\theta)\rangle_\theta \approx \omega(90°)$, gives angle-integrated spectrum as

$$\frac{d\sigma^1_{nnx}(\varepsilon,E_n)}{d\varepsilon} \approx \frac{d\tilde{\sigma}^1_{nnx}(\varepsilon,E_n)}{d\varepsilon} + \sqrt{\frac{\varepsilon}{E_n}}\frac{\langle\omega(\theta)\rangle_\theta}{E_n - \varepsilon}. \qquad (5)$$

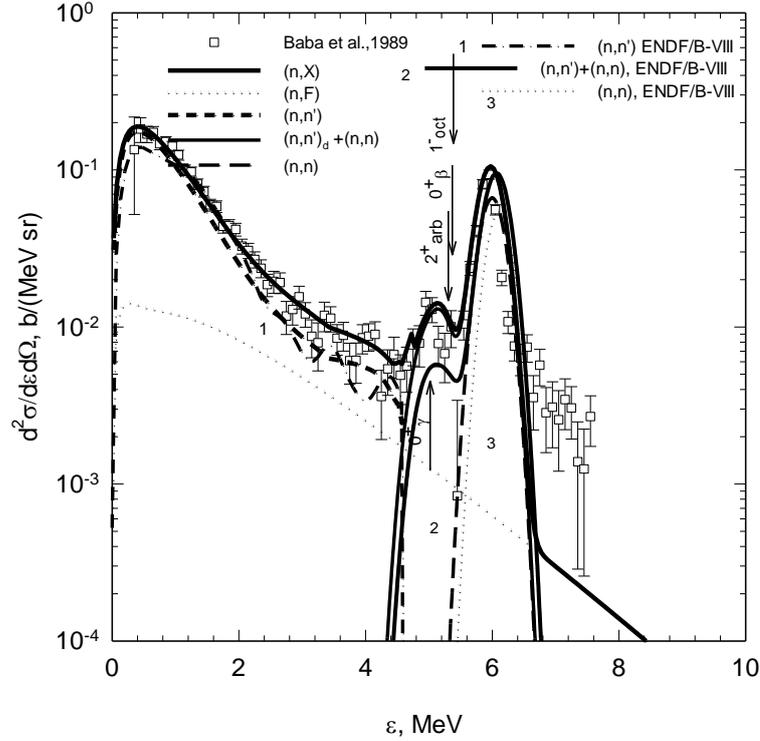

Fig. 2. Double differential neutron emission spectra for $^{232}$Th+$n$ at $E_n$ =6.1 MeV, $\theta \approx 120°$, and its partial constituents; full line – $(n,nX)$; dotted line – $(n,F)$; dashed line – $(n,n\gamma)^1$; dash-double-dotted line – $(n,2n)^1$; dashed line – $(n,2n)^2$; dash-dotted line–$(n,3n)^1$; dashed line – $(n,3n)^2$; full line – $(n,3n)^3$; dashed line – sum of $(n,n)_d$ and $(n,n\gamma)$ for discrete levels; ○ – [2, 3].

To retain the flux conservation in cross section and spectra calculations the compound reaction cross sections normalized to account for extra neutron emission:

$$\sigma_c(E_n) = \sigma_a(E_n)(1 - q - \tilde{q}), \qquad (6)$$

here, $q$–ratio of pre-equilibrium neutrons in a standard pre-equilibrium model [11], $\tilde{q}$ –value easily obtained using equation (5) [5]. The compound contribution to the emission spectrum of $(n,nX)^1$ reaction is

$$\frac{d\tilde{\sigma}^1_{nnx}(\varepsilon,E_n)}{d\varepsilon} = \sum_{J,\pi} W_A^{J\pi}(E_n - \varepsilon,\theta). \qquad (7)$$

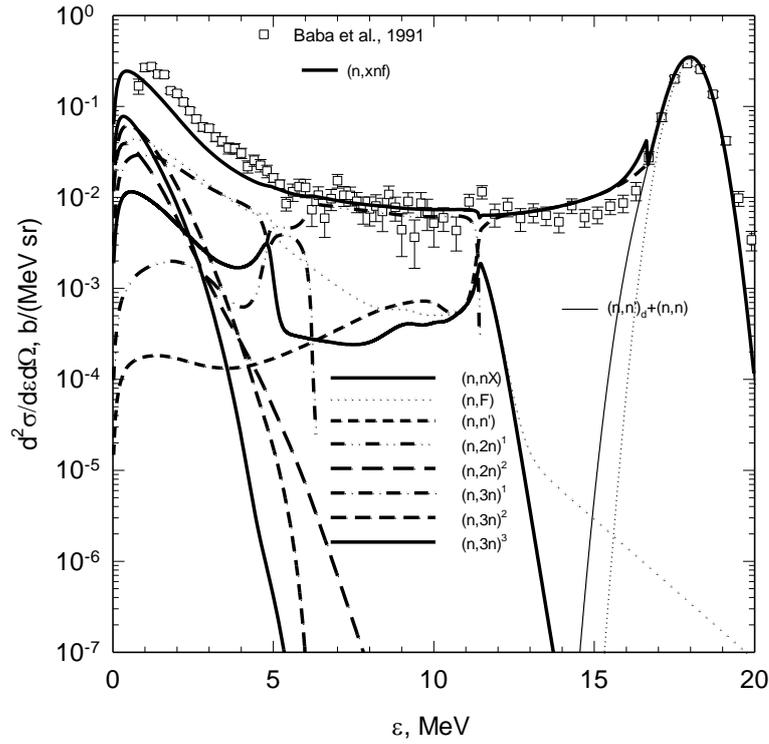

Fig. 3. Double differential neutron emission spectra for $^{232}$Th+$n$ at $E_n$ =18 MeV, $\theta \approx 30°$, and its partials; full line – $(n,nX)$; dotted line – $(n,F)$; dashed line – $(n,n\gamma)^1$; dash double dotted line –$(n,2n)^1$; dashed line– $(n,2n)^2$; dash–dotted line – $(n,3n)^1$; dashed line – $(n,3n)^2$; full line – $(n,3n)^3$; dashed line – $(n,n)$ + $(n,n\gamma)$ for discrete levels; ○ – [2, 3].

Population of states with spin/parity $J^\pi$ and excitation energy $U=E_n–\varepsilon$, after first neutron emission at angle $\theta$ depends on fission probability of $(A+1)$ nuclide. It defines the exclusive spectra of each partial reaction in STAPRE [11] framework, $W_A^{J\pi}(E_n - \varepsilon, \theta)$ is the population of excited states of residual nuclide $A$. The net effect of these procedures is the adequate approximation of double differential NES and angular distributions of $^{232}$Th$(n,nX)^1$ first neutron inelastic scattering in continuum, which corresponds to $U$= 1.2~6 MeV excitations for $E_n$ =1.16 MeV~20 MeV.

Angular anisotropy of NES of $^{232}$Th+$n$ interaction as observed in [2, 3] allows to extract the anisotropic contribution to double differential spectra of the first neutron, relevant for the excitations of first residual nuclide of 1.2~6 MeV and attribute observed NES anisotropy mostly to the component of $^{232}$Th$(n,n\gamma)^1$ reaction. The experimental quasi-differential emissive neutron spectra for $^{238}$U+$n$ interactions [14] revealed the inadequacy of NES modelling in [12, 13] and stimulated further efforts of NES modelling [15] aimed to abandon fictitious levels.

Figures 1 and 2 show NES at $E_n$ ~6.1 MeV for forward and backward scattering of first $(n,nX)^1$ neutron. The contribution of prompt fission neutrons to NES is exceptionally low. The inelastic scattering when residual nuclide excitation energy is larger than ~1.2 MeV and elastic scattering are the major contributors to the NES. The step-like structure to the left of elastic peak is due to direct excitation of collective levels of $\beta$–bands with $K^\pi =0^+$, $\gamma$–bands with $K^\pi =2^+$ and octupole band $K^\pi = 0^-$. Both elastic and inelastic scattering contributors to the NES are much dependent on angle $\theta$. The elastic and inelastic scattering contributors to the NES of ENDF/B-VIII [12, 13], though they roughly approximate the NES around elastic peak and inelastic scattering contribution when residual nuclide excitation energy is larger than ~1.2 MeV, waive off the direct excitation of $\beta$–bands with $K^\pi =0^+$, $\gamma$–bands with $K^\pi =2^+$ and octupole band $K^\pi =0^-$. The fictitious levels with $J^\pi = 2^+$, $3^-$ as in [12, 13] we avoid.

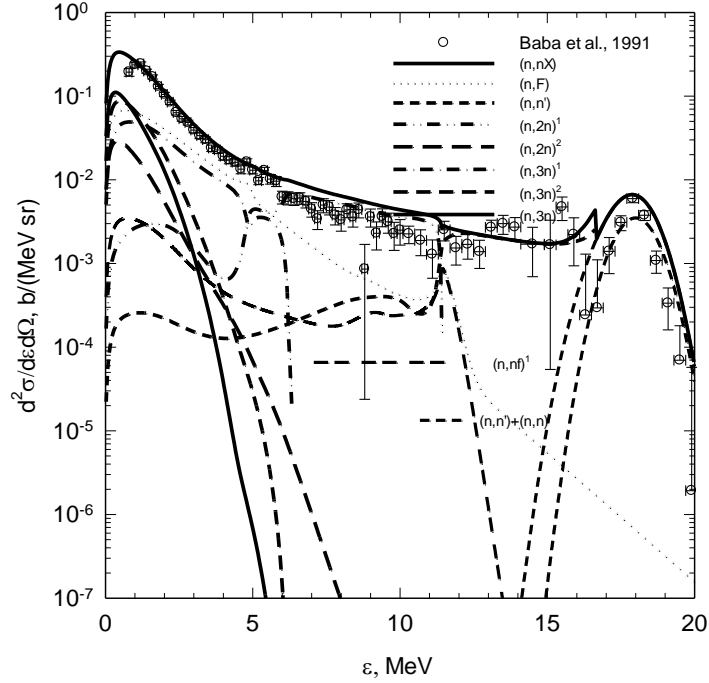

Fig. 4. Double differential neutron emission spectra for $^{232}$Th+$n$ at $E_n$ =18 MeV, $\theta \approx 120°$, and its partial contributions: full line – ($n,nX$); dotted line – ($n,F$); dashed line – ($n,n\gamma)^1$; dash-double-dotted line – $(n,2n)^1$; dashed line – $(n,2n)^2$; dash-dotted line–$(n,3n)^1$; dashed line – $(n,3n)^2$; full line – $(n,3n)^3$; dashed line – sum of ($n,n$) and ($n,n\gamma$) for discrete levels;○ – [2, 3].

PFNS anisotropy occurs because some portion of $(n,nX)^1$ neutrons might be involved in exclusive pre-fission neutron spectra. It was observed and interpreted in $^{238}$U($n,xnf$) [4], $^{235}$U($n,xnf$) [5, 6, 16, 17] and $^{239}$Pu($n,xnf$) [6, 16, 18–20] reactions. Since the contribution of PFNS to the NES of $^{232}$Th+$n$ interaction is relatively low, the major evidence of NES angular anisotropy would occur in ($n,n\gamma$) reaction. However, in $^{232}$Th($n,xnf$) reactions PFNS would demonstrate different response to forward and backward $(n,xnf)^1$ neutron emission relative to the incident neutron momentum, much stronger than in case of $^{238}$U($n,xnf$) reaction [4]. The partial PFNS components of $^{232}$Th($n,F$), shown on Fig. 1, were calculated as follows.

Prompt fission neutron spectra $S(\varepsilon, E_n, \theta)$ at angle $\theta$ relative to the incident neutron beam is a superposition of exclusive spectra of pre-fission neutrons, $(n,nf)^1$, $(n,2nf)^{1,2}$, $(n,3nf)^{1,2,3}$ –

$\dfrac{d^2\sigma_{nxn}^k(\varepsilon, E_n, \theta)}{d\varepsilon d\theta}$ ($x$=1, 2, 3; $k=1,...,x$), and spectra of neutrons, emitted by fission fragments, $S_{A+1-x}(\varepsilon, E_n, \theta)$:

$$S(\varepsilon, E_n, \theta) = \tilde{S}_{A+1}(\varepsilon, E_n, \theta) + \tilde{S}_A(\varepsilon, E_n, \theta) + \tilde{S}_{A-1}(\varepsilon, E_n, \theta) + \tilde{S}_{A-2}(\varepsilon, E_n, \theta) =$$
$$\nu_p^{-1}(E_n, \theta) \cdot \{ \nu_{p1}(E_n) \cdot \beta_1(E_n, \theta) S_{A+1}(\varepsilon, E_n, \theta) + \nu_{p2}(E_n - \langle E_{nnf}(\theta)\rangle) \beta_2(E_n, \theta) S_A(\varepsilon, E_n, \theta) +$$
$$+ \beta_2(E_n, \theta) \dfrac{d^2\sigma_{nnf}^1(\varepsilon, E_n, \theta)}{d\varepsilon d\varepsilon} + \nu_{p3}(E_n - B_n^A - \langle E_{n2nf}^1(\theta)\rangle - \langle E_{n2nf}^2(\theta)\rangle) \beta_3(E_n, \theta) S_{A-1}(\varepsilon, E_n, \theta) + \beta_3(E_n, \theta) \times$$
$$\left[ \dfrac{d^2\sigma_{n2nf}^1(\varepsilon, E_n, \theta)}{d\varepsilon d\theta} + \dfrac{d^2\sigma_{n2nf}^2(\varepsilon, E_n, \theta)}{d\varepsilon d\theta} \right] + \nu_{p4}(E_n - B_n^A - B_n^{A-1} - \langle E_{n3nf}^1(\theta)\rangle - \langle E_{n3nf}^2(\theta)\rangle - \langle E_{n3nf}^3(\theta)\rangle) \times$$
$$\beta_4(E_n, \theta) S_{A-2}(\varepsilon, E_n, \theta) + \beta_4(E_n, \theta) \left[ \dfrac{d^2\sigma_{n3nf}^1(\varepsilon, E_n, \theta)}{d\varepsilon d\theta} + \dfrac{d^2\sigma_{n3nf}^2(\varepsilon, E_n, \theta)}{d\varepsilon d\theta} + \dfrac{d^2\sigma_{n2nf}^3(\varepsilon, E_n, \theta)}{d\varepsilon d\theta} \right] \}. \quad (8)$$

In equation (8) $\tilde{S}_{A+1-x}(\varepsilon, E_n, \theta)$ is the contribution of $x$-chance fission to the observed PFNS $S(\varepsilon, E_n, \theta)$, $\langle E_{nxnf}^k(\theta)\rangle$ – average energy of $k$–th neutron of ($n,xnf$) reaction with spectra $\dfrac{d^2\sigma_{nxn}^k(\varepsilon, E_n, \theta)}{d\varepsilon d\theta}$, $k \leq x$. Spectra $S(\varepsilon, E_n, \theta)$, $S_{A+1-x}(\varepsilon, E_n, \theta)$ and $\dfrac{d^2\sigma_{nxn}^k(\varepsilon, E_n, \theta)}{d\varepsilon d\theta}$ are normalized to unity. Index $x$ denotes the fission chances $^{232}$Th($n,nxnf$) of $^{232}$Th($n,F$) after emission of $x$ pre-fission neutrons, $\beta_x(E_n, \theta) = \sigma_{n,xnf}(E_n, \theta)/\sigma_{n,F}(E_n, \theta)$ – contribution of $x$–th fission chance to the observed fission cross section, $\nu_p(E_n, \theta)$ is the observed average number of prompt fission neutrons, $\nu_{px}(E_{nx})$ – average number of prompt fission neutrons, emitted by fission fragments of $^{233-x}$Th nuclides. Spectra of prompt fission neutrons, emitted from fragments, $S_{A+2-x}(\varepsilon, E_n, \theta)$, as proposed in [21], were approximated by the sum of two Watt [22] distributions with different temperatures, the temperature of the light fragment being higher.

The Fig. 1 shows also partial contributions of $^{232}$Th($n,f$), $^{232}$Th($n,nf$) and $^{232}$Th($n,nf$)[1] to NES at $E_n$ ~6.1 MeV for forward scattering of first ($n,nX$)[1] neutron. The contribution of exclusive $^{232}$Th($n,nf$)[1] pre-fission neutrons to the NES is exceptionally low. With increase of the incident neutron energy $E_n$ up to ~20 MeV exclusive spectra of pre-fission neutrons, ($n,nf$)[1], ($n,2nf$)[1,2], ($n,3nf$)[1,2,3], when accounted for properly, will strongly influence the shapes of NES and PFNS [23].

Figures 3 and 4 show NES at $E_n$~18 MeV for forward and backward scattering, respectively. The contribution of prompt fission neutrons to the NES is still rather low, however the contribution of exclusive prefission neutron spectra is of peculiar shape. In an approach pursued in [12, 13] the direct excitation of levels other than those of ground state band is waived off and coupling strength erroneously transferred to fictitious levels with $J^\pi = 2^+, 3^-$. The inelastic scattering, when residual nuclide excitation energy is 1.2~6 MeV as well as elastic scattering are the major contributors to the NES at $\varepsilon > E_{nnf1}$, the latter being the cutoff energy of the prefission ($n,nf$)[1] neutron. The direct excitation of collective levels of $\beta$–bands with $K^\pi = 0^+$, $\gamma$–bands with $K^\pi = 2^+$ and octupole band $K^\pi = 0^-$ is no longer pronounced as a step-like structure to the left side of elastic peak. It leads just to broadening of quasi-elastic peak. In case of backward scattering the direct excitation is even more important contributor to NES because backward elastic scattering is suppressed. The anisotropic part of double differential spectra of first neutron of $(n,nX)^1$ reaction relevant for the excitation energy amounting fission barrier value of $^{232}$Th, will be strongly pronounced in exclusive spectra of $(n,nf)^1$, $(n,2nf)^1$ и $(n,2n)^1$ at $E_n >12$ MeV at various emission angles of first pre-fission neutron.

Modelling the angular distribution for the exclusive spectra of pre-fission neutrons of $^{235}$U$(n,xnf)^{1,...x}$ and $^{239}$Pu$(n,xnf)^{1,...x}$ we reproduced [5, 6] measured data of [17–20], namely, the ratios of $\langle S(\varepsilon, E_n, \Delta\theta)\rangle_{\Delta E_n} / \langle S(\varepsilon, E_n, \Delta\theta^1)\rangle_{\Delta E_n}$ at $\Delta\theta \approx 35^o - 40^o$, $\Delta\theta^1 \approx 130^o - 140^o$ and rather wide energy range of incident neutrons $\Delta E_n \sim 15$–17.5 MeV (see Fig. 5). Angular and spin correlations during prompt fission neutron emission are rather tedious, if possible at all, meanwhile the main factor for observed features of PFNS like $\langle S(\varepsilon, E_n, \Delta\theta)\rangle_{\Delta E_n} / \langle S(\varepsilon, E_n, \Delta\theta^1)\rangle_{\Delta E_n}$ and $\langle E(\theta \approx 37.5^o)\rangle / \langle E(\theta^1 \approx 135^o)\rangle$, is the excitation energy of fissioning nuclides emerging after emission of $x$ neutrons. The same approach might be pursued in case of $^{232}$Th$(n,xnf)^{1,...x}$ reactions.

The excitation energy of residual nuclides, after emission of $(n,xnf)$ neutrons, is decreased by the binding energy of emitted neutron $B_{nx}$ and its average kinetic energy:

$$U_x = E_n + B_n - \sum_{x, 1 \leq k \leq x} (\langle E^k_{nxnf}(\theta)\rangle + B_{nx}). \qquad (9)$$

Fission energy of $^{232}$Th$(n,F)$ reaction is distributed between fission fragments kinetic energy, their excitation energy and kinetic energy of pre-fission neutrons. The excitation energy of fission fragments is

$$E_{nx} = E_r - E^{pre}_{fx} + E_n + B_n - \sum_{x, 1 \leq k \leq x} (\langle E^k_{nxnf}(\theta)\rangle + B_{nx}). \qquad (10)$$

Value of TKE, kinetic energy of fission fragments prior prompt neutron emission, $E^{pre}_F$, is approximated as a superposition of partial TKE of $^{233-x}$Th nuclides as

$$E^{pre}_F(E_n) = \sum_{x=0}^{X} E^{pre}_{fx}(E_{nx}) \cdot \sigma_{n,xnf} / \sigma_{n,F}. \qquad (11)$$

Kinetic energy of fission fragments, i.e. post-fission fragments after neutron emission, $E^{post}_F$, are defined as

$$E^{post}_F \approx E^{pre}_F (1 - \nu_{post}/(A + 1 - \nu_{pre})). \qquad (12)$$

Similar relation was used for $E^{post}_f$ in [24] at $E_n < E_{nnf}$. Observed average number of prompt fission neutrons $\nu_p(E_n)$ is defined as

$$\nu_p(E_n) = \nu_{post} + \nu_{pre} = \sum_{x=1}^{X} \nu_{px}(E_{nx}) + \sum_{x=1}^{X} (x-1) \cdot \beta_x(E_n). \qquad (13)$$

The post-fission, $\nu_{post}(E_n)$, and pre-fission $\nu_{pre}(E_n)$ partials of $\nu_p(E_n)$ were obtained via consistent description of $\nu_p(E_n)$ and observed fission cross sections at $E_n <20$ MeV.

Contribution of $x$–th fission chance $(n,xnf)$ to the observed fission cross section $(n,F)$ is

$$\sigma_{nF}(E_n) = \sigma_{nf}(E_n) + \sum_{x=1}^{X} \sigma_{n,xnf}(E_n) \ . \tag{14}$$

That means the $(n,xnf)$ contributions are defined by fission probability $P_{f(A+1-x)}^{J\pi}(E)$ of $^{233-x}$Th nuclides:

$$\sigma_{n,xnf}(E_n) = \sum_{J\pi} \int_0^{U_x} W_{A+1-x}^{J\pi}(U) P_{f(A+1-x)}^{J\pi}(U) dU \ , \tag{15}$$

here $W_{A+1-x}^{J\pi}(U)$ –is the population of excited states of $(A+1-x)$ nuclides with excitation energy $U$ after emission of $x$ post-fission neutrons [25].

The direction of emission of $(n,nX)^1$ neutrons, as well as that of $(n,n\gamma)^1$, $(n,2n)^1$, $(n,3n)^1$ and $(n,nf)^1$, $(n,2nf)^1$ and $(n,3nf)^1$ neutrons, is correlated with the momentum of the incident neutrons. The direction of the neutrons emitted from the fission fragments correlates with the fission axis direction mostly. Pre-fission neutrons influence the PFNS shape of $^{232}$Th$(n,F)$ in the energy range of $E_n \sim E_{nnf1} \div 20$ MeV. They influence also the shape of TKE of fission fragments and products [26], prompt neutron number, mass distributions and produce the step-like structures in observed fission cross section of rather peculiar shape. The variation of observed average energies $\langle E \rangle$ in the vicinity of $^{238}$U$(n,xnf)$ reaction thresholds, as shown in [23, 27, 28], is defined by the exclusive spectra of $(n,xnf)^{1...x}$ neutrons. The amplitude of variations of $\langle E \rangle$ in case of $^{238}$U$(n,F)$ [23, 27–29] was confirmed by PFNS measured data $^{238}$U$(n,F)$ [30] in $E_n \sim 1$–20 MeV energy range.

Henceforth omitted are the indexes $J^\pi$ in fission, $\Gamma_f$, neutron $\Gamma_n$ and total $\Gamma$ widths described in [31], as well as relevant summations. The angular dependence of partial width, calculated with spin and parity conservation, is due to dependence of excitation energy of residual nuclides on emission angle of first $(n,nX)^1$ neutron. The exclusive spectra of pre-fission $(n,nf)^1$ neutron is

$$\frac{d^2\sigma_{nnf}^1(\varepsilon, E_n, \theta)}{d\varepsilon d\theta} = \frac{d^2\sigma_{nnx}^1(\varepsilon, E_n, \theta)}{d\varepsilon d\theta} \frac{\Gamma_f^A(E_n - \varepsilon, \theta)}{\Gamma^A(E_n - \varepsilon, \theta)}. \tag{16}$$

First neutron spectra of $(n,2nf)^1$ for reaction $(n,2nf)$, is defined as:

$$\frac{d^2\sigma_{n2nf}^1(\varepsilon, E_n, \theta)}{d\varepsilon d\theta} = \int_0^{E - B_n^A} \frac{d^2\sigma_{n2nx}^1(\varepsilon, E_n, \theta)}{d\varepsilon d\theta} \frac{\Gamma_f^{A-1}(E_n - B_n^A - \varepsilon - \varepsilon_1)}{\Gamma^{A-1}(E_n - B_n^A - \varepsilon - \varepsilon_1)} d\varepsilon_1 \tag{17}$$

here first neutron spectra of $(n,2nx)$ reaction, i.e. $(n,2nx)^1$, is defined by the neutron spectrum of $(n,nX)^1$ and neutron emission probability of nuclide $A$ as:

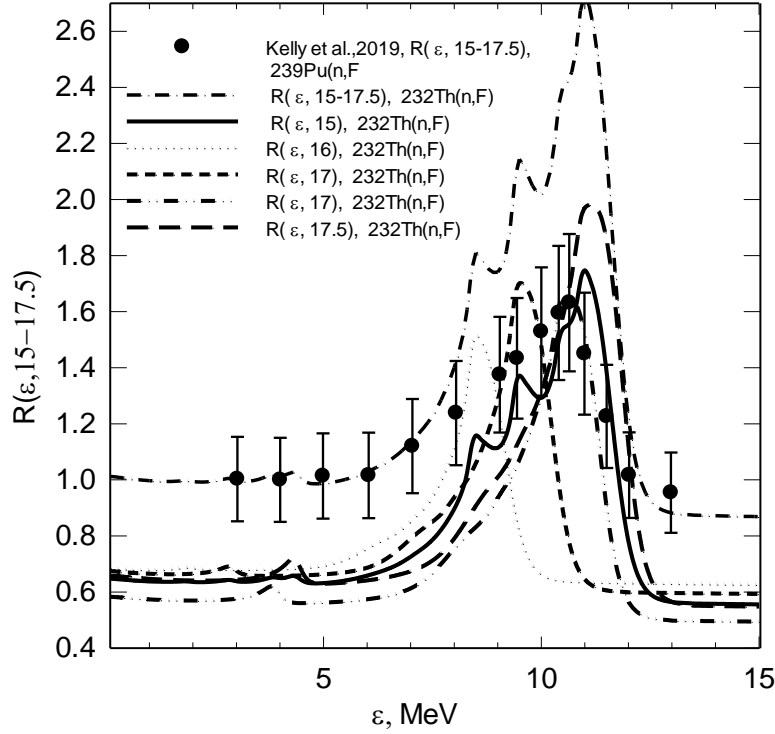

Fig. 5 Measured ratios $R^{exp} = S(\varepsilon, E_n \approx 15-17.5, \Delta\theta)/S(\varepsilon, E_n \approx 15-17.5, \Delta\theta^1)$ of $^{239}$Pu$(n,F)$ PFNS and calculated $R(\varepsilon, 15 \div 17.5)$ for "forward", $\Delta\theta \sim 35°–40°$ and "backward" emission; $\Delta\theta^1 = 130°–140°$; ●– $^{232}$Th$(n,F)$ [1]; full line – $^{232}$Th$(n,F)$ PFNS normalized to unity; dashed line – $^{232}$Th$(n,F)$ PFNS equated at $\varepsilon \sim 3–5$ MeV; dash-dotted line – $^{232}$Th$(n,F)$ PFNS equated at $\varepsilon \sim 3–5$ MeV; dotted line – partials of $^{232}$Th$(n,F)$ $R(\varepsilon, 15 \div 17.5)$ at $E_n \sim 15$ MeV, $E_n \sim 16$ MeV, $E_n \sim 17$ MeV and $E_n \sim 17.5$ MeV.

$$\frac{d^2\sigma^1_{n2nx}(\varepsilon, E_n, \theta)}{d\varepsilon d\theta} = \frac{d^2\sigma^1_{nnx}(\varepsilon, E_n, \theta)}{d\varepsilon d\theta} \frac{\Gamma^A_n(E_n - \varepsilon, \theta)}{\Gamma^A(E_n - \varepsilon, \theta)}. \quad (18)$$

Spectra of first and next neutrons of $^{232}$Th$(n,3nf)$ reaction are covered in [23], but their contribution is quite low. Adopted phenomenological approach enables to reproduce NES in case of $^{232}$Th+n interactions at $E_n \sim 1–18$ MeV. Exclusive pre-fission neutron spectra of $^{232}$Th$(n,xnf)^{1,2}$ are shown on Figs. 3 and 4 as $\dfrac{\sigma_{n,xnf}(E_n, \theta)}{4\pi} \dfrac{d\sigma^{1,2}_{nnf}(\varepsilon, E_n, \theta)}{d\varepsilon}$ at angles $\theta \sim 30°$ и $\theta \sim 135°$. They comprise small part of $(n,nX)^1$ spectrum, but relatively large part of the observed PFNS.

Angular distributions of $^{239}$Pu$(n,xnf)$ pre-fission neutrons at $E_n \sim 14–18$ MeV, measured in [18], were quite well described as $\sim 0.25\,\omega(\theta)$, if $\theta > 135°$, then $0.25\,\omega(\theta = 135°)$. Estimate of pre-fission neutrons contribution in [18] they extracted as difference of observed PFNS and some approximated estimate of post-fission neutrons evaporated from fission fragments.

Angular anisotropy of PFNS relative to incident neutron beam was detected in $^{239}$Pu$(n,F)$ [1] at $E_n$~15–17.5 MeV range, $\Delta\theta$ ~35°–40° (forward direction) and $\Delta\theta^{1}$=130°–140° (backward direction) ranges. The data normalization obtained by equating observed PFNS at $\varepsilon$~3–5 MeV energy range. Alternative representation of PFNS, against that shown on Fig.3 in [18], as a ratio $R^{exp} = S(\varepsilon, E_n \approx 15-17.5, \Delta\theta)/S(\varepsilon, E_n \approx 15-17.5, \Delta\theta^{1})$ for $\Delta\theta$ ~35°–40° (forward direction) and $\Delta\theta^{1}$=130°–140° (backward scattering) is virtually independent upon the normalizations adopted in [18].

On Fig. 5 $R^{exp}$ of $^{239}$Pu$(n,F)$ at $E_n$ ~15—17.5 MeV $\Delta\theta$ ~35°–40° (forward direction) and $\Delta\theta^{1}$=130°–140° (backward direction) compared with of $^{232}$Th$(n,F)$ calculated ratio

$$R(\varepsilon, 15 \div 17.5) \approx \frac{\int_{15}^{17.5} \nu_p(E_n, \approx 30^{o})\sigma_{nF}(E_n, \approx 30^{o})S(\varepsilon, E_n, \theta \approx 30^{o})\varphi(E_n)dE_n}{\int_{15}^{17.5} \nu_p(E_n, \theta \approx 135^{o})\sigma_{nF}(E_n, \theta \approx 135^{o})S(\varepsilon, E_n, \theta \approx 135^{o})\varphi(E_n)dE_n}, \qquad (19)$$

here $\varphi(E_n)$ is the incident neutron spectrum, which is not known. Spectra $S(\varepsilon, E_n, \theta)$ normalized to unity. As a first order approximation $R(\varepsilon, 15 \div 17.5)$ (14) might be calculated as a ratio of lumped sums for $E_n$ ~ 15 MeV, $E_n$ ~ 16 MeV, $E_n$ ~ 17 MeV and $E_n$ ~ 17.5 MeV $\nu_p(E_n,\theta)\sigma_{nF}(E_n,\theta)S(\varepsilon, E_n \approx 15-17.5, \Delta\theta)$ and $\nu_p(E_n,\theta)\sigma_{nF}(E_n,\theta)S(\varepsilon, E_n 5-17.5, \Delta\theta^{1})$. Values of $\nu_p(E_n,\theta)$ and $\sigma_{nF}(E_n,\theta)$ calculated at the same $E_n$ and $\theta$ as those in $S(\varepsilon, E_n \approx 15-17.5, \Delta\theta)$. In case of angular dependent observables for $^{232}$Th$(n,F)$ hidden structures in lumped $R(\varepsilon, 15 \div 17.5)$ constituents (evident for monochromatic beams) are smoothed, then $R^{exp}$ and calculated $R(\varepsilon, 15-17.5)$ seem to have similar shapes, but $R(\varepsilon, 15-17.5)$ is shifted downwards. Smooth line of $R(\varepsilon, 15-17.5)$ on Fig. 5 obtained by assuming in equation (19) that numerator and denominator values at $\varepsilon$~3–5 MeV energy range equal, as assumed in [18]. In case of $^{232}$Th$(n,F)$ and $^{239}$Pu$(n,F)$ both $R^{exp}$ and $R(\varepsilon, 15-17.5)$ are less then unity at $\varepsilon > E_{nnf1}$, that might be due to influence of angular dependence of $(n,xnf)$ neutron emission on the fission chances distribution.

The calculated anisotropy of pre-fission neutrons of $^{232}$Th$(n,xnf)$ reaction is appreciably higher than in case of $^{239}$Pu$(n,F)$. That is due to correlation of anisotropy of pre-fission neutrons with contribution of emissive fission reaction $^{232}$Th $(n,nf)$ to the observed fission cross section $^{232}$Th $(n,F)$, PFNS and angular anisotropy of NES. In case of $^{232}$Th$(n,F)$ and $^{239}$Pu$(n,F)$ at $\varepsilon > E_{nnf1}$, both $R^{exp}$ and calculated $R(\varepsilon, 15-17.5)$ are less then unity, that also is due to influence of angular dependence of $(n,xnf)$ neutron emission on the fission chances distribution.

Angular dependence of the first pre-fission neutron in reactions $(n,nf)^{1}$ and $(n,2nf)^{1}$ allows to interpret the experimental data trend observed in case of ratio of average energies for "forward" and "backward" emission of pre-fission neutrons in $^{235}$U$(n,xnf)^{1,2,3}$ [17] and $^{239}$Pu$(n,xnf)^{1,2,3}$ [18] reactions. The ratio of $\langle E(\theta) \rangle / \langle E(\theta^{1}) \rangle$ in case of $^{232}$Th$(n,F)$ for "forward", $\Delta\theta$ ~35°–40° and "backward", $\Delta\theta^{1}$=130°–140°, pre-fission neutron emission steeply increases starting from $E_n \sim 10$–12 MeV. The angular anisotropy of $(n,xnf)^1$ neutrons emission is due to pre-equilibrium/semidirect emission of first neutron in $(n,nX)^1$ (see Fig. 6).

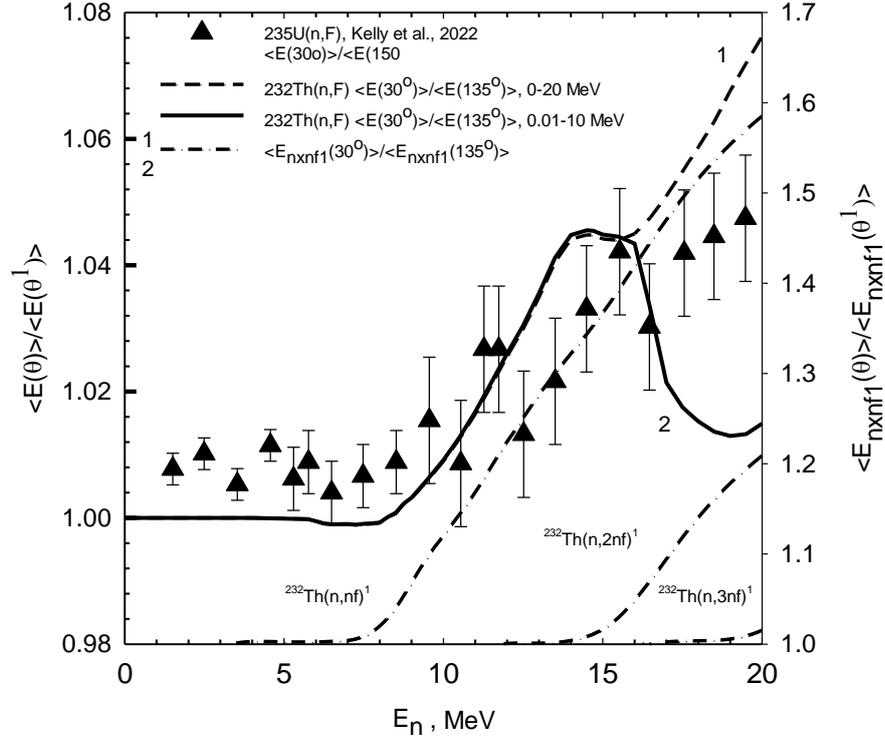

Fig. 6. Ratio of average energies of $^{235}$U$(n,F)$ PFNS $\langle E(\theta)\rangle/\langle E(\theta^1)\rangle$: ▲ – $\langle E(\theta \approx 30^o)\rangle/\langle E(\theta^1 \approx 135^o)\rangle$, $\varepsilon \sim 1$–12 MeV [17]; and $^{232}$Th$(n,F)$ PFNS $\langle E(\theta)\rangle/\langle E(\theta^1)\rangle$: dashed line $-\langle E(\theta \approx 30^o)\rangle/\langle E(\theta^1 \approx 135^o)\rangle$, $\varepsilon \sim 1$–20 MeV; full line $-\langle E(\theta \approx 30^o)\rangle/\langle E(\theta^1 \approx 135^o)\rangle$, $\varepsilon \sim 0.89$–10 MeV; dash–dotted lines 1, 2, 3 $-\langle E_{n,xnf}(\theta \approx 30^o)\rangle/\langle E_{n,xnf}(\theta^1 \approx 135^o)\rangle$, $x$=1, 2, 3.

The ratio of average energies of exclusive neutron spectra of $^{232}$Th$(n,nf)^1$, $\dfrac{d^2\sigma_{nnf}^1(\varepsilon, E_n, \theta \approx 30^o)}{d\varepsilon d\theta}$ and $\dfrac{d^2\sigma_{nnf}^1(\varepsilon, E_n, \theta \approx 135^o)}{d\varepsilon d\theta}$, $\langle E_{n,xnf}(\theta \approx 30^o)\rangle/\langle E_{n,xnf}(\theta^1 \approx 135^o)\rangle$, is much higher than that of $\langle E(\theta)\rangle/\langle E(\theta^1)\rangle$, however it follows the shape of experimental ratio $\langle E(\theta \approx 30^o)\rangle/\langle E(\theta^1 \approx 135^o)\rangle$[17] of $^{235}$U$(n,F)$. Angular dependence of the ratio of average energies of exclusive neutron spectra $^{232}$Th$(n,2nf)^1$: $\dfrac{d^2\sigma_{n2nf}^1(\varepsilon, E_n, \theta \approx 30^o)}{d\varepsilon d\theta}$ and $\dfrac{d^2\sigma_{n2nf}^1(\varepsilon, E_n, \theta \approx 150^o)}{d\varepsilon d\theta}$ is much weaker. In the ratio of average energies of exclusive neutron spectra of $^{232}$Th$(n,3nf)^1$, $\dfrac{d^2\sigma_{n3nf}^1(\varepsilon, E_n, \theta \approx 30^o)}{d\varepsilon d\theta}$ and $\dfrac{d^2\sigma_{n3nf}^1(\varepsilon, E_n, \theta \approx 150^o)}{d\varepsilon d\theta}$, the angular dependence is quite weak.

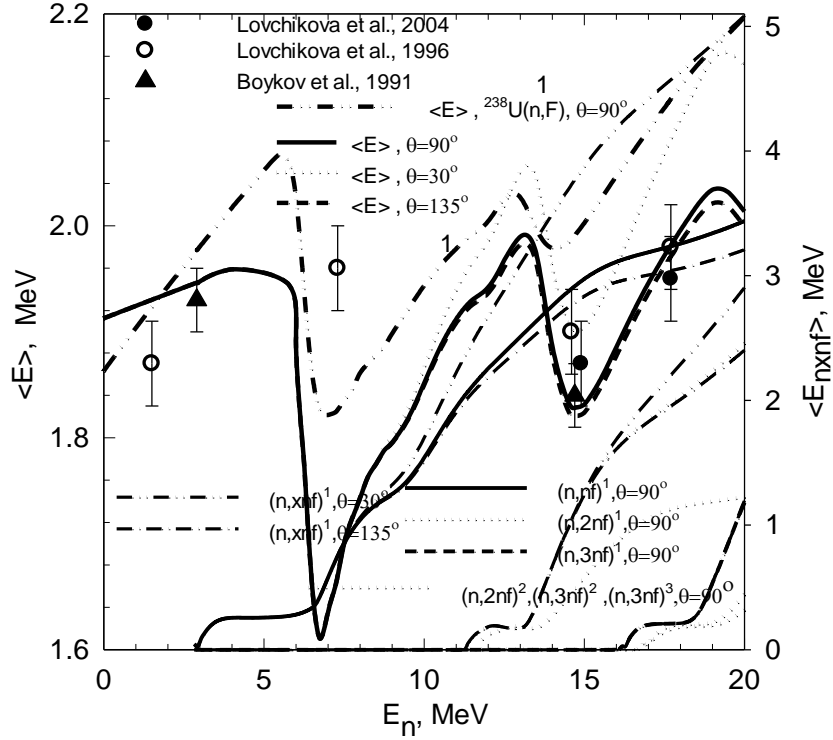

Fig. 7 Average energies of PFNS $\langle E \rangle$ of $^{232}$Th$(n,F)$ and $^{238}$U$(n,F)$: ● –[32]; ○–[33]; ▲ – [34]; full line – $\langle E(90^o)\rangle$ $^{232}$Th$(n,F)$; dotted line – $\langle E(30^o)\rangle$; dashed line–$\langle E(135^o)\rangle$; full line – $\langle E_{n,nf}(\theta \approx 90^o)\rangle$; dotted line – $\langle E_{n,2nf}(\theta \approx 90^o)\rangle$; dashed line – $\langle E_{n,3nf}(\theta \approx 90^o)\rangle$; dotted line – $\langle E_{n,xnf}(\theta \approx 90^o)\rangle$ of $(n,2nf)^2$ and $(n,3nf)^{2,3}$; dash–dotted line –$\langle E_{n,xnf}(\theta \approx 135^o)\rangle$; dash–double dotted line –$\langle E_{n,xnf}(\theta \approx 30^o)\rangle$; dash–double dotted line, 1 – $\langle E \rangle$ $^{238}$U$(n,F)$.

Ratios $\langle E(\theta \approx 30^o)\rangle/\langle E(\theta^1 \approx 135^o)\rangle$ are virtually independent upon the lower threshold of neutron detection, while the dependence upon angular range and value of higher neutron detection threshold ($\varepsilon$~10, $\varepsilon$~12 or $\varepsilon$~20 MeV) is crucial. For emitted neutrons energy range of $\varepsilon$~1–12 MeV or $\varepsilon$~0–20 MeV, as evidenced on Fig. 6, shape of $\langle E(\theta \approx 30^o)\rangle/\langle E(\theta^1 \approx 135^o)\rangle$ of $^{232}$Th$(n,F)$ is roughly consistent with measured data for $^{235}$U$(n,F)$ up to $E_n$ ~16 MeV. For exclusive neutron spectra of $^{232}$Th$(n,nf)^1$ the ratios of $\dfrac{d^2\sigma^1_{nnf}(\varepsilon, E_n, \theta \approx 30^o)}{d\varepsilon d\theta}$ and $\dfrac{d^2\sigma^1_{nnf}(\varepsilon, E_n, \theta \approx 135^o)}{d\varepsilon d\theta}$ average energies are also much higher than those of $\langle E(\theta)\rangle/\langle E(\theta^1)\rangle$, but their shape is virtually consistent with that of $\langle E(\theta \approx 30^o)\rangle/\langle E(\theta^1 \approx 135^o)\rangle$ [17] (see Fig. 6). Average energy $\langle E \rangle$ is a rough integral signature of PFNS, however the angular anisotropy of pre-fission neutron emission exerts quite an influence on its values. Dependence of $\langle E \rangle (E_n)$ in case of $^{232}$Th$(n,F)$ is compared with measured data [32–34] on Fig. 7. The estimates of $\langle E \rangle$ for PFNS of $^{232}$Th$(n,F)$ are strongly correlated with PFNS shape. The influence of exclusive neutron spectra of $(n,nf)^1$ and $(n,2nf)^{1,2}$ which they exert on $\langle E \rangle$ in case of $^{232}$Th$(n,F)$ are much stronger than in case of $^{238}$U$(n,F)$ [23, 27–30]. Drop in $\langle E \rangle (E_n)$ in the vicinity of $^{232}$Th$(n,2nf)$ reaction threshold is the deepest ever observed in measured PFNS data.

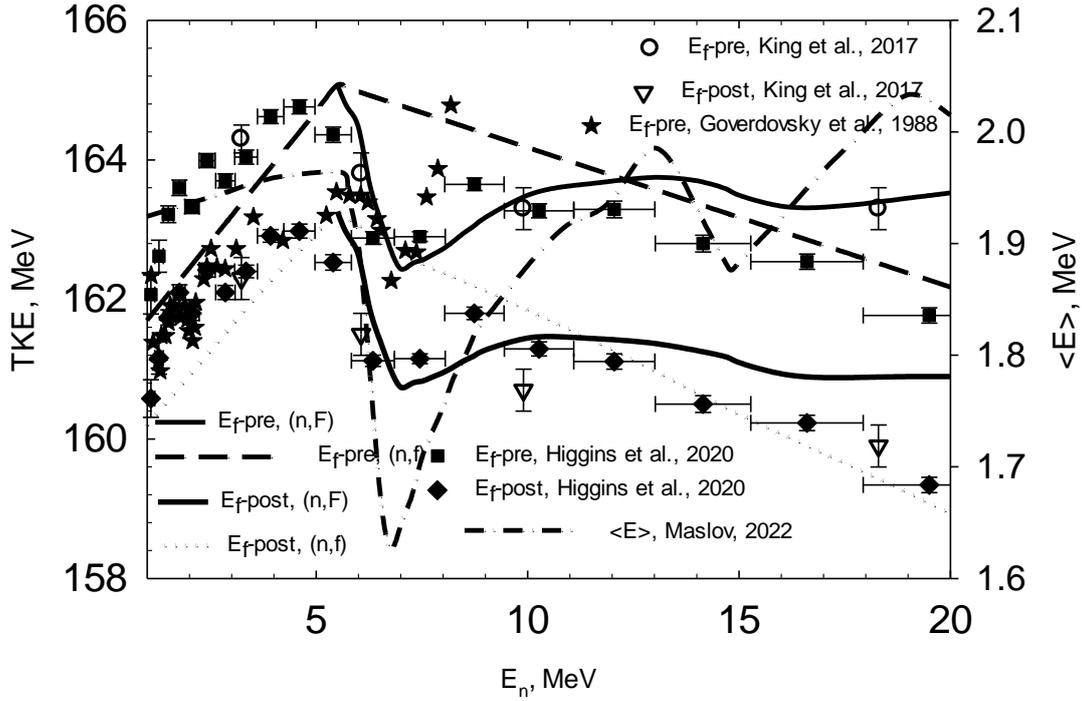

Fig. 8 Average total kinetic energies TKE of $^{232}$Th$(n, F)$: black stars–[35]; ● –[36]; ■ – [36]; ♦ –[37]; ■ – [37]; dash–dotted line – $\langle E \rangle$ of PFNS.

Another complication of observables in neutron-induced fission of $^{232}$Th is the total kinetic energy TKE trend. Local minimum in TKE for the pre-neutron emission fission fragments in $^{232}$Th$(n,F)$ around $^{232}$Th$(n,nf)$ threshold first observed in [35]. That strong TKE variation is due to the pre-fission $(n,xnf)$ neutrons. Contribution of the $(n,xnf)$ reaction to the $\sigma_{n,F}$ of $^{232}$Th$(n,F)$ around $E_n \sim 7$ MeV is exceptionally high [1], as well as provoked dip in TKE [26]. Partial contributions of $(n,xnf)$, initially fixed in [23, 25], reproduce TKE variations. TKE values $E_f^{pre}$ ($E_f^{post}$) before (after) prompt neutron emission from fission fragments were calculated with equations (11–14) provided above. Components $\nu_{post}$ and $\nu_{pre}$ of $\nu_p$ are defined via $\nu_p$ and PFNS analysis at $E_n$ up to 20 MeV. Assuming $E_f^{pre}(E_n)$ for $^{233-x}$Th nuclides are similar to that of $^{233}$Th, we obtained TKE both before and after prompt fission neutron emission, as shown on the Fig. 8. Neutrons emitted from the fission fragments $\nu_{post}$ and composite nuclide $^{233}$Th, $\nu_{pre}$, are predicted. Calculated TKE values shown on the Fig. 8 are consistent with the observed $^{232}$Th+$n$ data on neutron cross sections and PFNS. Straight lines approximate TKE values for the first chance fission of $^{233}$Th nuclide. The (n,xnf)-neutrons influence TKE values $E_f^{pre}$ and $E_f^{post}$, it is pronounced in case of $^{232}$Th(n,F) reaction mostly as a sharp drop around $^{232}$Th(n,nf) reaction threshold. That is due to the transition states structure of $^{232}$Th fissioning nuclide and competition of $^{232}$Th(n,nγ) and $^{232}$Th(n,nf) reactions at $E_n$ ≤6.5 MeV and $^{232}$Th(n,2n) at higher energies. In case of $^{238}$U(n,F) TKE behaves in a mirror-like character [38], in [29] the observed local maxima were interpreted in similar fashion. The major difference in case of first chance fission of $^{233}$Th or $^{239}$U is, respectively, the increasing and decreasing trend of TKE.

Analysis of neutron emission spectra $^{232}$Th+n and prompt fission neutron spectra of $^{232}$Th(n,F) evidence correlations of many observed data structures with (n,xnf)$^{1...x}$ pre-fission neutrons. In case of NES observed angular anisotropy is due to angular dependence of elastic scattering, direct excitation of collective levels and pre-equilibrium emission of (n,nX)$^1$ neutrons. Proper description of $^{232}$Th+n NES is attained when ground state band levels $J^\pi = 0^+, 2^+, 4^+, 6^+, 8^+$ are coupled within rigid rotator model, while those of $\beta$–bands with $K^\pi = 0^+$, $\gamma$–bands with $K^\pi = 2^+$ and octupole band $K^\pi = 0^-$ are coupled within soft deformable rotator model. NES of $^{232}$Th+n at $E_n$ ~6, ~12, ~14, ~18 MeV described. The net effect of these procedures for $E_n$ <20 MeV is the adequate approximation of angular distributions of $^{232}$Th(n,nX)$^1$ first neutron inelastic scattering in continuum, which corresponds to $U$=1~6 MeV excitations of $^{232}$Th.

Pre-fission neutron spectra turned out to be quite soft as compared with neutrons emitted by excited fission fragments. The net outcome of that is the decrease of $\langle E \rangle$ in the vicinity of the $^{232}$Th(n,xnf) thresholds of $^{232}$Th(n,F). The amplitude of the $\langle E \rangle$ variation is much higher in case of $^{232}$Th(n,F) as compared with $^{238}$U(n,F). The correlation of PFNS shape with different angles of emission of (n,xnf)$^1$ neutrons and emissive fission contributions for $^{232}$Th(n,F) is established. The angular anisotropy of exclusive pre-fission neutron spectra strongly influences the PFNS shapes and $\langle E \rangle$. These peculiarities are due to strong emissive fission contributions in $^{232}$Th(n,F). Predicted ratio of PFNS $\langle E \rangle$ for "forward" and "backward" emission of pre-fission neutrons seems to be the largest among stable actinide target nuclides. It steeply increases alongside with the increase of the average energies of the exclusive pre-fission neutron spectra [39].

**References**
1. V. M. Maslov, M. Baba, A. Hasegawa, A. B. Kagalenko, N.V. Kornilov, N.A. Tetereva, INDC(BLR)-16, IAEA, Vienna (2003), https://www-nds.iaea.org/publications/indc/indc-blr-0016/.
2. M. Baba, H. Wakabayashi, N. Ito et al., JAERI-M-89-143, 1989.
3. S. Matsuyama, M. Baba, N. Ito et al., JAERI-M-91-032, 219, 1991.
4. V.M., Maslov LXXII International Conference " NUCLEUS-2022, Fundamental problems and applications", Moscow, July, 11—16, 2022, Book of Abstracts, p.168, https://events.sinp.msu.ru /event/ 8/attachments/181/875 nucleus-2022-book-of-abstracts-www.pdf.
5. V.M. Maslov, Physics of Particles and Nuclei Letters, 20, 1401 (2023).
6. V.M. Maslov, Yad. Fyz., 86, 562 (2023).
7. V.M. Maslov, Yu.V. Porodzinskij, M. Baba, A. Hasegawa, Bull. RAS, Ser. Fyz., 67, 1597 (2003).
8. V.M. Maslov, Yu.V. Porodzinskij, N.A. Tetereva et al., Nucl. Phys. A, 764, 212 (2006).

9. V. M. Maslov, M. Baba, A. Hasegawa, A. B. Kagalenko, N.V. Kornilov, N.A. Tetereva, INDC(BLR)-14, IAEA, Vienna (2003), https://www-nds.iaea.org/publications/indc/indc-blr-0014/.
10. M. Dupuis, S. Hilaire, S. Peru, EPJ Web of Conferences, 146, 12002 (2017).
11. M. Uhl and B. Strohmaier, IRK-76/01, IRK, Vienna, 1976.
12. P.G. Young, M. Chadwick, R. MacFarlane et al., Nucl. Data Sheets, 108, 2589 (2007).
13. D. A. Brown, M. B. Chadwick, R. Capote et al., Nucl. Data Sheets 148, 1 (2018).
14. A.M. Daskalakis, R.M.Bahran, E.J. Blain et al., Ann. of Nucl. Energy 73, 455 (2014).
15. M. R. Mumpower, D. Neudecker, H. Sasaki, et al., Phys. Rev. C, 107, 034606 (2023).
16. V.M. Maslov In: Proc. LXXII Intern. Conf. Nucleus 2022, Fundamental problems and applications, Moscow, 11—16 July, 2022, Book of abstracts, p. 111, https://events.sinp.msu.ru/event/8/attachments/181/875 nucleus–2022–book–of–abstracts–www.pdf.
17. K. J. Kelly, J.A. Gomez, M. Devlin et al, Phys. Rev. C 105, 044615 (2022).
18. K. J. Kelly, T. Kawano, J.M. O'Donnel et al., Phys. Rev. Lett., 122, 072503 (2019).
19. P. Marini, J. Taieb, B. Laurent et al., Phys. Rev. C, 101, 044614 (2020).
20. K. J. Kelly, M. Devlin, O'Donnel J.M. et al., Phys. Rev. C, 102, 034615 (2020).
21. N.V. Kornilov, A.B. Kagalenko, F.-J. Hambsch, Yad. Fiz., 62, 209 (1999).
22. B.E. Watt, Phys. Rev., 87, 1037, (1952).
23. V.M. Maslov, Yu. V. Porodzinskij, M. Baba, A. Hasegawa, N.V. Kornilov, A.B. Kagalenko and N.A. Tetereva, Phys. Rev. C, 69, 034607 (2004).
24. D. Madland, Nucl. Phys. A 772, 113 (2006).
25. V.M. Maslov. Nucl. Phys. A743, 236 (2004).
26. V.M. Maslov 27[th] International Seminar on Interactions of Neutrons with Nuclei, 2020, May, Dubna, Russia, http://isinn.jinr.ru/past-isinns/isinn-27/abstracts/Maslov.pdf/
27. V.M. Maslov, Yad. Fiz., 71, 11 (2008).
28. V. M. Maslov, EPJ Web Conf. 8, 02002 (2010); https://epjwoc.epj.org/articles/epjconf/abs/2010/07/epjconf_efnudat2010_02002/epjconf_efnudat2010_02002.html
29. V.M. Maslov, Physics of Particles and Nuclei Letters, 20, 565 (2023); https://www.pleiades.online/cgi-perl/search.pl?type=abstract&name=physpnlt&number=4&year=23&page=565.
30. K. J. Kelly, M. Devlin, J.M. O'Donnel et al., Phys. Rev. C, 108, 024603 (2023).
31. V.M. Maslov, Phys. Rev. C, 72, 044607 (2005).
32. G.N.Lovchikova,A.M.Trufanov,M.I.Svirin,V.A.Vinogradov Yad. Fyz. 67, 914 (2004); https://link.springer.com/article/10.1134/1.1777281
33. G.N. Lovchikova, A.M. Trufanov, VANT, Ser. Yadernye Konstanty, 1, 102, 1996; INDC(CCP)-409,115,1997.
34. G.S. Boykov, V.D. Dmitriev, G.A. Kydyaev, Ostapenko, M.I. Svirin, G.N. Smirernkin, Yad. Fyz., 53, 628 (1991).
35. A.A. Goverdovsky, B.D. Kuzminov, V.F. Mitrofanov, A.I. Sergachev, Proc. Conf. on Nucl. Data for Sci. and Technol., Mito. 1988 p. 695.
36. J. King, W. Loveland, J. S. Barrett et al., Eur. Phys. Journ. A, 53, 238 (2017).
37. D. Higgins, U. Greife, F. Tovesson et al., Phys. Rev. C, 101, 014601 (2020).
38. C. Zoller, Ph.D. thesis, TH Darmstadt, 1995, http://www-win.gsi. de/charms/data.htm.
39. V.M. Maslov, Proc. 28[th] International Seminar on Interactions of Neutrons with Nuclei, 2021, May, 24-28, Dubna, Russia, Book of Abstracts, p. 113, http://isinn.jinr.ru/past-sinns/isinn 28/ISINN 28 %20 Abstract%20 Book.pdf.